\documentclass[final,5p,times,twocolumn]{elsarticle}

\usepackage{lineno}
\usepackage[hidelinks]{hyperref}

\usepackage{graphicx}
\usepackage{amsmath} 
\usepackage{siunitx}
\let\svthefootnote\thefootnote
\newcommand\freefootnote[1]{%
  \let\thefootnote\relax%
  \footnotetext{#1}%
  \let\thefootnote\svthefootnote%
}
\def\pmbanner{{\hrule height 1 pt}\vskip35pt{NIMA POST-PROCESS BANNER TO BE REMOVED AFTER FINAL ACCEPTANCE}\vskip35pt{\hrule height 4pt}\vskip20pt}

\begin{document}

\begin{frontmatter}

\title{\pmbanner Characterisation of analogue MAPS fabricated in 65 nm technology for the ALICE ITS3}

\author[1,2]{Kunal Gautam\corref{cor1}\fnref{fn1}
  }
\ead{kunal.gautam@cern.ch}

\author[3]{Ajit Kumar
}


 \cortext[cor1]{Corresponding author}
 \fntext[fn1]{On behalf of the ALICE Collaboration}

\affiliation[1]{organization={Vrije Universiteit Brussel}, 
                 addressline={Pleinlaan 2},
                 postcode={1050}, 
                 city={Brussels}, 
                 country={BE}}
\affiliation[2]{organization={Universität Zürich}, 
                 addressline={Winterthurerstr. 190},
                 postcode={8057}, 
                 city={Zürich}, 
                 country={CH}}
\affiliation[3]{organization={Université de Strasbourg}, 
                 addressline={CNRS, IPHC UMR 7178},
                 postcode={F‐67000}, 
                 city={Strasbourg}, 
                 country={FR}}

\begin{abstract}
The ALICE ITS3 project foresees the use of ultra-light MAPS, developed in the 65 nm imaging process, for the vertex detector in the ALICE experiment at the LHC to drastically improve the vertexing performance. This new development, initiated by an international consortium of the ALICE ITS3 collaboration and the CERN EP R\&D project, enhances the overall MAPS performance.

Small-scale prototypes are designed to study the analogue properties of the TPSCo 65 nm technology and compare the charge collection performance in different processes, pitches, pixel geometries, and irradiation levels. Recent results from lab and test-beam characterisation detailing the efficiency and the spatial resolution of the APTS with different pixel geometries and pitches satisfy the ALICE ITS3 requirements. A quantitative evolution of the charge collection and sharing among pixels is evident in the CE-65 with different in-pixel readouts. Attaining a spatial resolution better than 3 µm with a 10 µm pitch and over $99\%$ efficiency in the moderate irradiation environment of ALICE also supports the viability of using 65 nm MAPS for FCC-ee vertex detectors.
\end{abstract}


\end{frontmatter}


\section{ALICE ITS3 and the Small-Scale Prototypes}
\label{sec:intro}
The Inner Tracking System upgrade (ITS3) \cite{The:2890181} of the ALICE detector foresees the use of Monolithic Active Pixel Sensors (MAPS) developed in the 65 nm imaging process. The ITS3 upgrade is expected to drastically reduce the material budget and improve the pointing resolution by a factor of 2 over the entire momentum range of interest. Process modifications in the form of a blanket low-dose n-type implant and gaps at the edges of pixel boundaries are introduced to achieve full depletion of the sensor and a more uniform electric field while keeping the effective pixel capacitance low \cite{SNOEYS201790,Munker_2019}. These modifications speed up charge collection and reduce charge sharing among neighbouring pixels. Two small-scale analogue prototypes were designed to characterise the charge-collection properties of the 65 nm process and study the effects of the modification.

The Analogue Pixel Test Structure (APTS) \cite{rinella2024characterisation} is a small $6\times6$ pixel matrix with a fast direct analogue readout of the central $4\times4$ pixels, and the Circuit Exploratoire 65 (CE-65) \cite{BUGIEL2022167213} features a $64\times32$ or $48\times32$ pixel matrix with a rolling-shutter analogue readout. The APTS has a source-follower in-pixel readout configuration, while the CE-65 hosts three sub-matrices with either an AC/DC-coupled preamplifier or source-follower in-pixel readout configuration. Both prototypes are designed in various pitch sizes ranging from \SI{10}{\micro\meter} to \SI{25}{\micro\meter}.

\section{Lab and Test-beam Measurements}
\label{sec:results}
An extensive lab and test-beam campaign was performed to assess the performance of the chips by measuring their charge-collection properties, detection efficiencies, and energy and position resolutions.

\subsection{Energy Calibration}
\label{subsec:calibration}
\begin{figure}[t]
\centering
\includegraphics[width=0.42\textwidth]{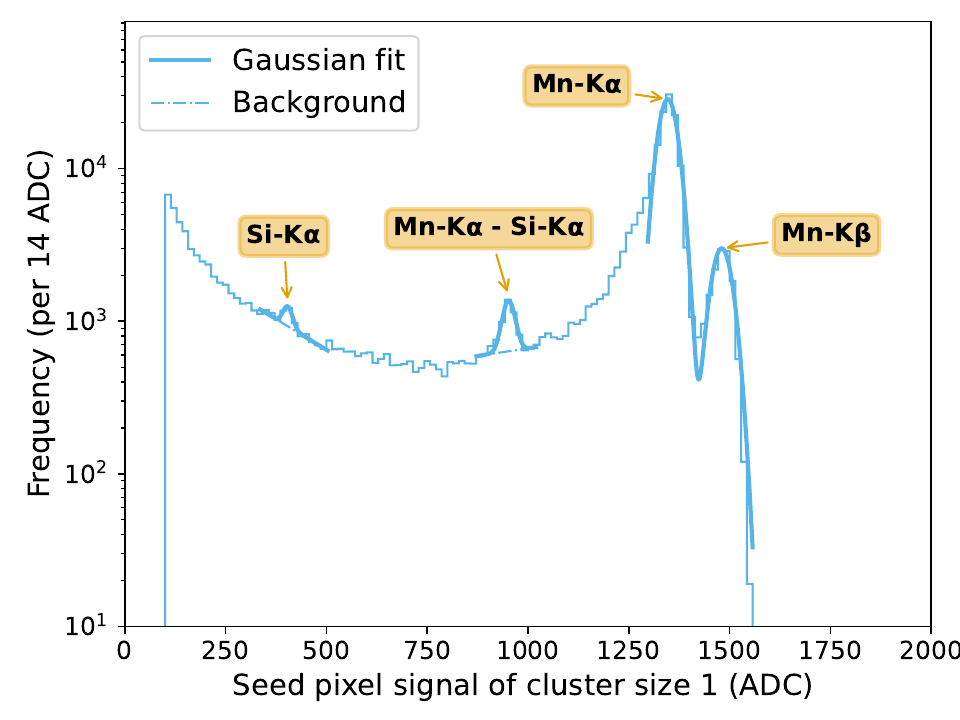}
\caption{${}^{55}\text{Fe}$ spectrum in single-pixel cluster events measured in the central 4 pixels of the APTS.}\label{fig:fe55}
\end{figure}

The energy calibration of the pixel response was performed with the ${}^{55}\text{Fe}$ radioactive source. The ${}^{55}\text{Fe}$ spectrum for events with cluster size of 1 is depicted in Figure \ref{fig:fe55}. A cluster is defined as the set of adjacent pixels within a $3\times3$ matrix centred around the seed pixel, the pixel collecting the highest charge above a threshold. The most prominent Mn-K$_{\alpha}$ peak is used to convert the ADC units to $e^-$. The linearity of the pixel response was established by fitting all peaks in the spectrum in a high-statistics scan.

\subsection{Detection Efficiency}
The test-beam measurements were performed with a telescope using 6 ALPIDE chips \cite{AGLIERIRINELLA2017583} as the reference planes and an APTS (Digital Pixel Test Structure) chip as the trigger with the device-under-test being APTS (CE-65).

Adding a low dose n-type implant slightly increases the noise RMS of the devices due to the increase in the effective capacitance of the pixels but also significantly improves the range of operation over $99\%$ efficiency. The considerable charge sharing among neighbouring pixels in the standard process causes a significant drop in efficiency with increasing thresholds compared to sensors with process modification. Different pixel geometries, in the form of varying shapes and sizes of the collection electrode and the PWELL enclosure, do not significantly affect the performance, the most notable being the variant with a larger collection electrode, which has a higher noise but is slightly more radiation tolerant.

\begin{figure}[h]
\centering
\includegraphics[width=0.5\textwidth]{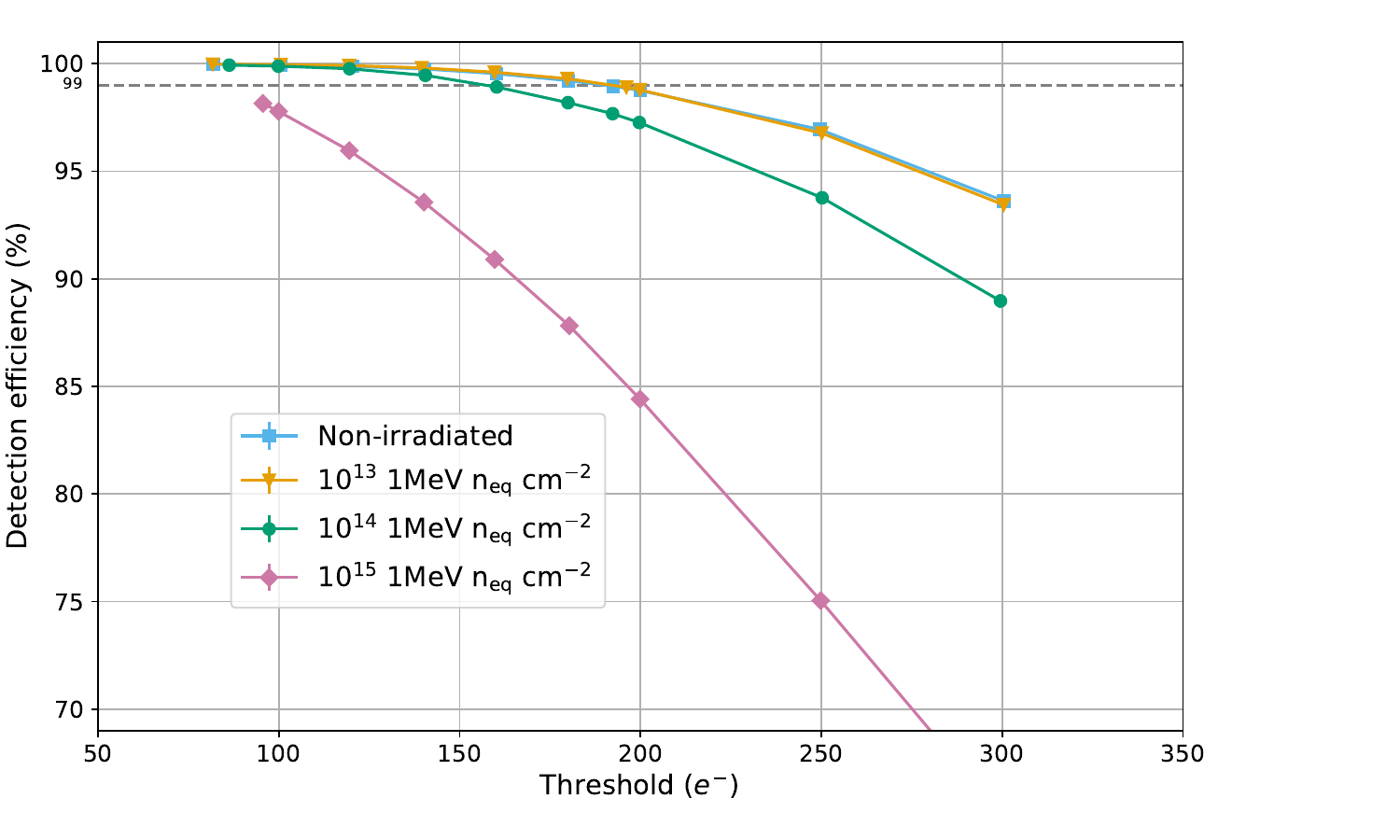}
\caption{Efficiency comparison of APTS irradiated to different levels of non-ionising radiation. The results correspond to sensors designed in the modified process with gap with the pitch size of \SI{20}{\micro\meter} and the bias voltage of 1.2 V. All threshold points are greater than $3\times$RMS of the noise.}\label{fig:irradiation}
\end{figure}

\begin{figure}[ht]
\centering
\includegraphics[width=0.49\textwidth]{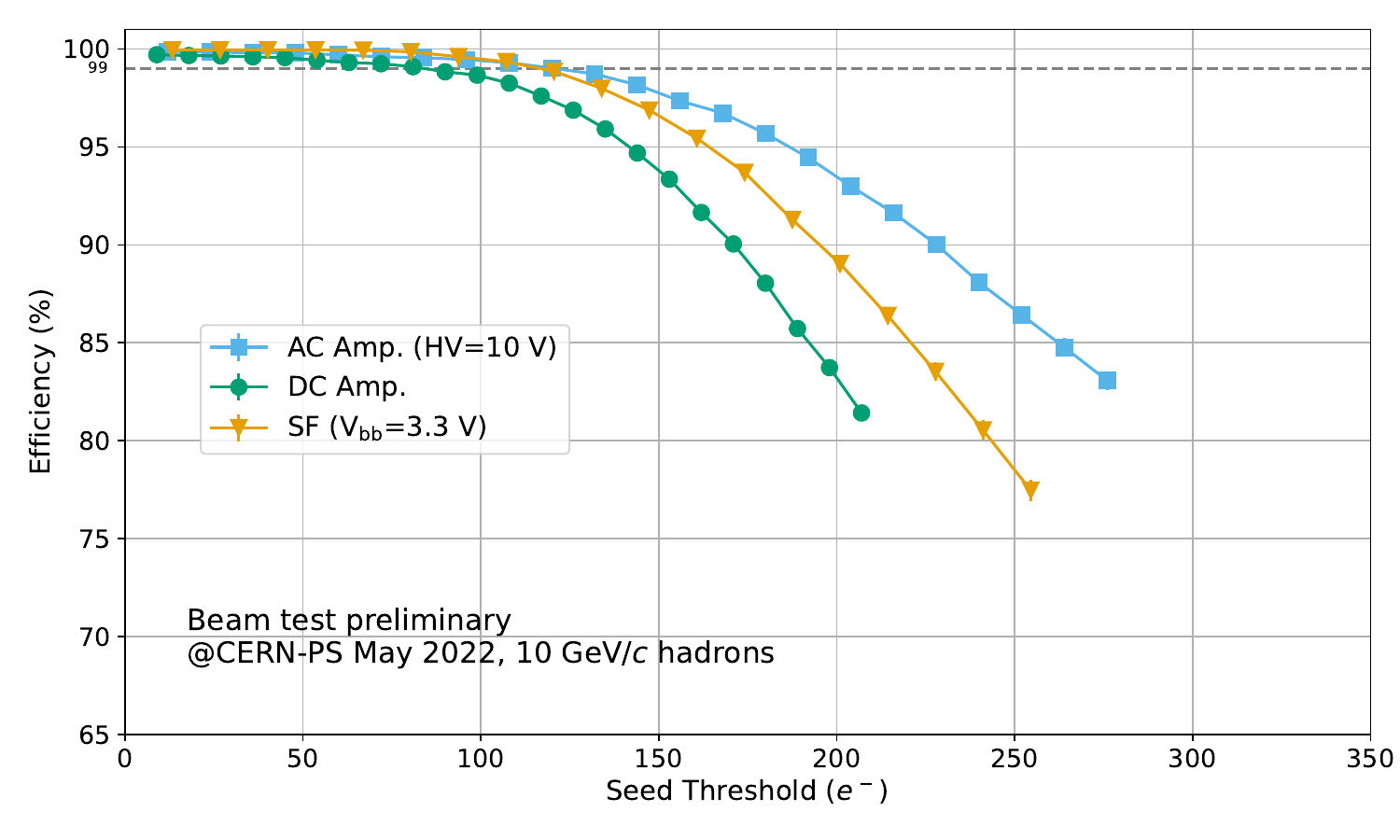}
\caption{Effect of depletion on detection efficiency through different in-pixel circuitry in the standard process CE-65.}\label{fig:readout}
\end{figure}

The APTS is performant up to irradiation levels of $10^{14}$ 1 MeV $\text{n}_{\text{eq}}~\text{cm}^{-2}$ with the application of the minimal bias voltage of 1.2 V, as is shown in Figure \ref{fig:irradiation}. The effect of depletion on the detection efficiency is depicted in Figure \ref{fig:readout}. An AC-coupled high voltage provides full depletion of the sensor while disentangled from the readout electronics, gaining high efficiencies even at high thresholds. The depletion depends on the bias voltage for the source-follower configuration, whereas the DC-coupled variant only achieves partial depletion in the absence of biasing, leading to reduced efficiencies at high thresholds.

\subsection{Spatial Resolution}
\begin{figure*}[ht]
\centering
\includegraphics[width=0.67\textwidth]{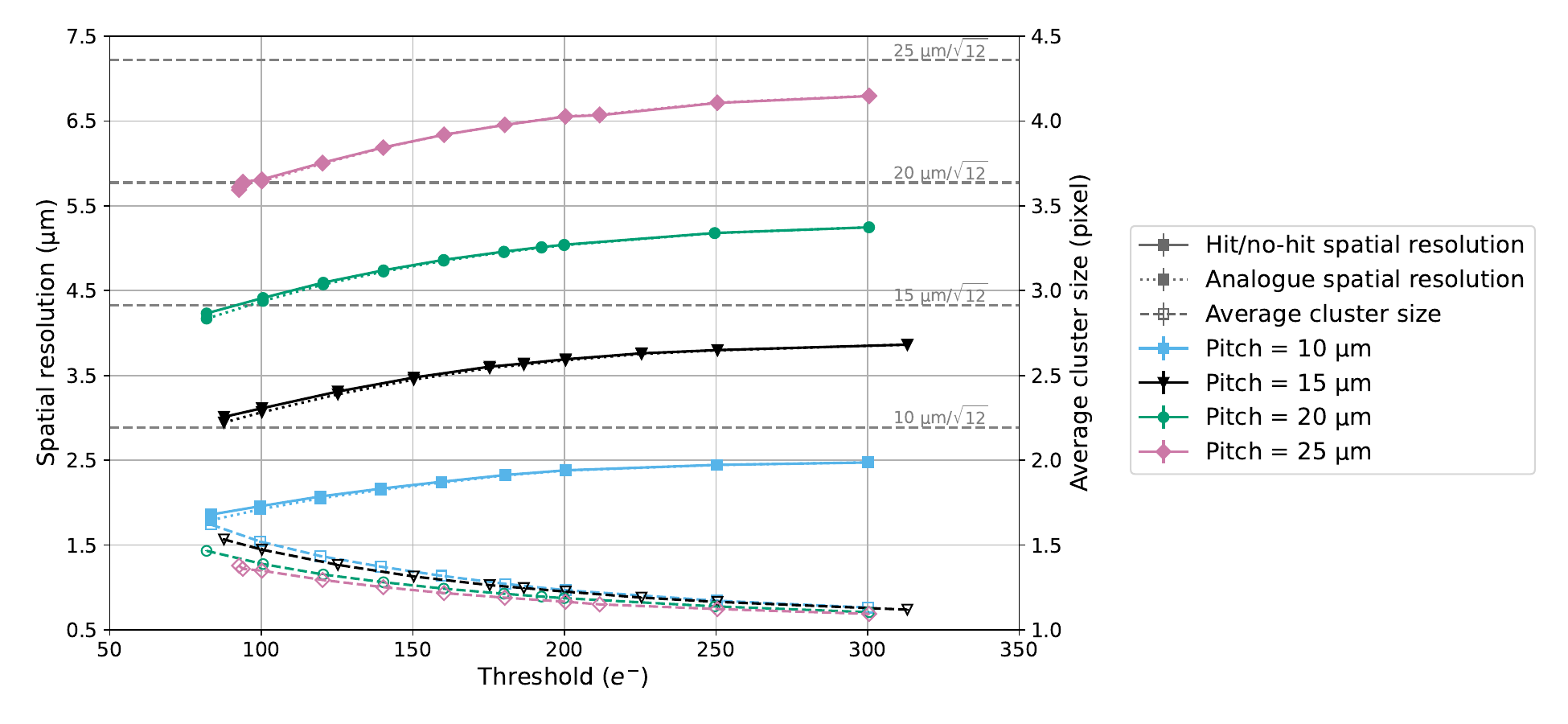}
\caption{Hit/no-hit resolution (solid lines), analogue resolution (dotted lines), and average cluster size (dashed lines) for several pitch sizes in modified-with-gap process APTS at the bias voltage of 1.2 V.}\label{fig:resolution}
\end{figure*}

The process modification reduces charge sharing among neighbouring pixels but does not eliminate it. Figure \ref{fig:resolution} shows that the achieved resolution is better than binary resolution with the average cluster size being greater than one. The APTS reaches a hit/no-hit resolution better than \SI{3}{\micro\meter} with a pitch size of \SI{10}{\micro\meter} while still operating with over $99\%$ efficiency. The sensors with larger pitch sizes in the modified process with gap show a bigger operational margin with efficiency over $99\%$.

\section{Conclusion and Outlook}
Additional tests are being pursued to study the effect of collection-electrode placement in the matrix on charge sharing. Further optimisation between detection efficiency and charge sharing is being investigated using pixel designs with AC-coupled high voltage for depletion. The excellent detection efficiency in moderate radiation environments and low spatial resolution make MAPS developed in 65 nm process a viable candidate for the vertex detectors at future lepton colliders, like FCC-ee.

The 65 nm CMOS process has now been validated, and the analogue properties of the sensors have been characterised by consistent results from the small and large matrix prototypes. The performance satisfies the specifications set forth by the ALICE ITS3 project.

\freefootnote{This project has received funding from the European Union’s Horizon 2020 Research and Innovation Programme under Grant Agreements 101004761 (AIDAinnova) and 101057511 (EURO-LABS).}

\bibliographystyle{elsarticle-num-names} 
\bibliography{refs}

\end{document}